\documentclass[aip,jcp,amsmath,amssymb,twocolumn,reprint]{revtex4}
\usepackage{hyperref}
\usepackage{natbib}
\usepackage{graphicx, color, soul}

\newcommand{\be}{\begin{equation}}
\newcommand{\ee}{\end{equation}}
\newcommand{\bea}{\begin{eqnarray}}
\newcommand{\eea}{\end{eqnarray}}

\newcommand{\kbt}{k_{\text{B}}T}

\newcommand{\Eq}[1]{Eq.~\eqref{#1}}

\newcommand{\Fig}{Fig. \ref}

\def\<{\left\langle}                 
\def\>{\right\rangle}                 

\begin{document} 
\title{Anomalous quantum and isotope effects in water clusters: Physical phenomenon, model artifact, or
  bad approximation?}  
\author{Sandra E. Brown}
\author{Vladimir A. Mandelshtam}
\affiliation{Department of Chemistry,
University of California, Irvine, CA 92697, USA}

\begin{abstract}
Free energy differences $\Delta F:=F-F_{\text{prism}}$ are computed for
several isomers of water hexamer relative to the ``prism'' isomer using the 
self-consistent phonons method.  
We consider the isotope effect defined by the quantity $\delta
F_{D_2O}:=\Delta F_{\rm D_2O}-\Delta F_{\rm H_2O}$, and the quantum
effect, $\delta F_{\hbar=0}:=\Delta F_{\hbar=0}-\Delta F_{\rm H_2O}$,
and evaluate them using different flexible water models. While both
$\delta F_{D_2O}$ and $\delta F_{\hbar=0}$ are found to be rather small 
for all of the potentials, they are especially small for two of the empirical models, 
\mbox{q-TIP4P/F} and \mbox{TTM3-F}, compared to \mbox{q-SPC/Fw} and the two
{\it ab~initio}-based models, WHBB and \mbox{HBB2-pol}.  
This qualitative difference in the properties of different water models cannot be
explained by one being ``more accurate" than the other.  
We speculate as to whether the observed anomalies are caused by the special properties
of water systems, or are an artifact of either the potential energy surface
form/parametrization or the numerical approximation used.
\end{abstract}

\maketitle

Changes in the properties of water upon isotopic substitution provide clear evidence of the large influence of nuclear quantum effects (e.g. zero-point motion, tunneling, etc.) on the behavior of water.
An oft-cited example is the shift in the melting  point of water by 3.8~K in deuterated water, and 4.5~K in tritiated water \cite{CRCHandbook2004}.
Trends such as these have long suggested that nuclear quantum effects destabilize the hydrogen bond network, ``softening" the structure of liquid water \cite{Miller2005, MorroneCar2008}.  
However, with the advent of improved water models, it has become increasingly clear that the situation is more complicated than may have been expected.  
Adding to the intrigue and complexity of the problem is the apparent competition between quantum effects at play in hydrogen-bonded systems \cite{Michaelides2011}, and in water in particular \cite{Habershon2009, Markland2012}, which can be tipped one way or the other depending upon the temperature \cite{Michaelides2011, Markland2012}.  
Quantum fluctuations of the intramolecular bond stretching tend to strengthen hydrogen bonds as a result of larger monomer dipole moments, while intermolecular quantum fluctuations tend to weaken them \cite{Habershon2009, Michaelides2011}.  
Compared to the extensive body of work on quantum effects in bulk water systems, far fewer studies exist examining isotope effects in the water hexamer cluster, in spite of its recognition as a representative species for three-dimensional hydrogen-bonded networks.  
The water hexamer is the smallest water cluster for which three-dimensional minimum energy structures are observed, as opposed to the two-dimensional ring structures favored by smaller clusters, earning it the endearing distinction of ``smallest drop of water" and status as an important benchmark for understanding the structure and dynamics of water \cite{GregoryClary1996, Xantheas2002,  Perez2012, SaykallyWales2012}.  
Aside from being an intrinsically interesting physical problem, there is seemingly endless motivation for developing a complete understanding of nuclear quantum effects in water, as isotope effects are relevant even in the context of many large-scale biological and environmental processes \cite{PaesaniVoth2008, Hoefs2009}.

A large number of water models have been proposed in the past. However, partly due to 
numerical reasons (to exclude the fast intra-molecular degrees of
freedom) most of those were based on ``rigid'' water molecules.  
(The interested reader is referred to the review \cite{vega2011}.)  
However, based on the previous studies \cite{Habershon2009, Markland2012}
we believe that a proper treatment of isotope effect must consider a flexible water model.  
When working with empirical models a well-known concern is how well the model can reproduce 
properties of water which were not used to parameterize it.  
As far as we are aware, hardly any of the existing empirical models can be used to accurately predict 
any specific property of water, unless this very property was actually used to design the model. 
At the same time, the current status of the electronic structure methods does not appear to
provide a sufficient level of accuracy for an {\it ab~initio} potential energy surface (PES) to
have predictive power. 
Moreover, even if we assume such a model to be sufficiently accurate, 
the cost associated with the evaluation of a high-quality {\it ab~initio} 
PES usually makes the inclusion of the quantum effects in 
a nuclear dynamics simulation too expensive.  
In this context it is important to mention recent efforts in the
construction of accurate parametric fits of high-level {\it ab~initio} water PES's, such as 
the WHBB \cite{Wang2011} and \mbox{HBB2-pol} \cite{Medders2013} potentials.  
In both WHBB and \mbox{HBB2-pol}, the monomer contribution is based on the same
Partridge-Schwenke fit \cite{PartridgeSchwenke1997} used in the \mbox{TTM3-F} model.
Most importantly, both WHBB and HBB2-pol include three-body terms (i.e., terms that explicitly couple coordinates of any three water monomers), which are not present in the other three potentials. In addition, they are both permutationally invariant in the sense that they allow for proton exchange.  
While the \mbox{HBB2-po}l PES is substantially faster and reportedly more accurate than WHBB \cite{Medders2013}, 
it is still much more expensive than most of the popular
empirical models, including \mbox{q-SPC/Fw} \cite{PaesaniVoth2006}, \mbox{q-TIP4P/F} \cite{Habershon2009}, and \mbox{TTM3-F} \cite{Fanourgakis2008}.  Although it was not included in the present study, we note the 
recent development of \mbox{MB-pol}, a ``first-principles" water potential which has been shown to be more accurate than \mbox{HBB2-pol} \cite{Babin2013_MB, Babin2014, Medders2014}. 

In a recent paper \cite{Babin2013} four of the above mentioned PES's
have been tested numerically using  replica exchange path integral
molecular dynamics (\mbox{RE-PIMD}) simulations, specifically, by computing the
free energy differences for several isomers of water hexamer. 
The convergence of these \mbox{RE-PIMD} results is being questioned elsewhere \cite{comment}.  
However, regardless of how accurate they are and regardless of 
how accurate any particular water model is, our present focus is not so 
much on establishing the absolute  ``truth'' on whether or not this or any 
other model correctly predicts that the cage isomer is energetically more 
favorable at low temperatures than the prism isomer, or vice versa.
Rather, our primary goal is to reveal possible generic properties of these water models which 
may be determined by the manner in which they have been constructed and parametrized.
 
In this study, in addition to the four PES's considered in
Ref.~\onlinecite{Babin2013}, we also consider \mbox{q-SPC/Fw} \cite{PaesaniVoth2006}, 
whose principal difference from \mbox{q-TIP4P/F} is the purely harmonic treatment of the OH-stretch.  
According to Habershon {\it et al} \cite{Habershon2009} this difference is responsible
for the significant difference in the magnitude of the isotope effect
between the two water models. 
 
All three empirical models, \mbox{q-SPC/Fw}, \mbox{q-TIP4P/F}  and  \mbox{TTM3-F}, unlike the two
{\it ab~initio}-based PES's, are not permutationally invariant and include only 
two-body interactions (i.e., interactions between not more than two water monomers), except for the fact that \mbox{TTM3-F} is polarizable.  
That is, by construction, the intramolecular forces in these models hold each hydrogen atom nearby
its oxygen, while both WHBB  and \mbox{HBB2-pol} allow, in principle, for
hydrogen exchange (albeit hardly correctly), thus leading to stronger 
couplings between the intramolecular and intermolecular degrees of freedom. 
With this said, it is not clear what the degree of quasi-separability
should be, and whether it is feasible to obtain a numerically
practical parametrization that combines it correctly with the permutational
invariance constraint and a correct description of hydrogen exchange.

The self-consistent phonons (SCP) method was first proposed several decades ago as a means to 
include anharmonic effects in the approximate treatment of the nuclear
dynamics of condensed phase systems \cite{Koehler1966, Gillis1968}.  
Interest in SCP in the context of finite systems has emerged only recently.  
In Ref.~\onlinecite{Calvo2010}  it was used to compute the fundamental frequencies of
aromatic hydrocarbons, and in Refs.~\onlinecite{Georgescu2011} and \onlinecite{Georgescu2012} 
for computing the ground states of  very large Lennard-Jones clusters.  
Given a quantum many-body system localized in an energy minimum at thermal equilibrium, 
SCP maps it to a reference harmonic system by optimizing the
free energy in the framework of the Gibbs-Bogolyubov variational principle. 
Recently we have shown how to overcome the numerical bottleneck of the method,
the accurate evaluation of multidimensional Gaussian averages of the potential 
and its derivatives \cite{Brown2013}.  
This was done by implementing quasi-Monte Carlo integration which exhibits a superior 
quasi-linear scaling with respect to the number of Monte Carlo points $N_{\rm MC}$, 
compared to the $\sqrt{N_{\rm MC}}$ scaling of standard Monte Carlo integration.  
For the systems mentioned here, the incorporation of quasi-Monte Carlo in SCP has been found
to reduce the computational cost of the method by several orders of magnitude, 
expanding its applicability range to {\it ab~initio} potentials.  

Though one can expect the accuracy of the SCP method to be significantly greater 
than that of the standard harmonic approximation, the SCP method is itself still an approximation.
The accuracy of SCP has been investigated in Ref.~\cite{Georgescu2013} and Ref.~\cite{comment} 
for the very case of water hexamer.  
In Ref.~\cite{Georgescu2013} the calculation of the isomer energy differences at zero temperature 
(i.e. for the isomer ground states), for which SCP turned out to be 
accurate. 
In Ref.~\cite{comment} we demonstrate that the classical free energy differences (i.e., $\Delta F_{\hbar=0}$) 
estimated by SCP agree well with those computed using a variant of reversible scaling, 
an exact-in-principal method \cite{KoningAntonelliYip1999, KoningCaiAntonelliYip2000}.  

\begin{figure*}
	\begin{centering} 
	\includegraphics[width=1.0\linewidth]{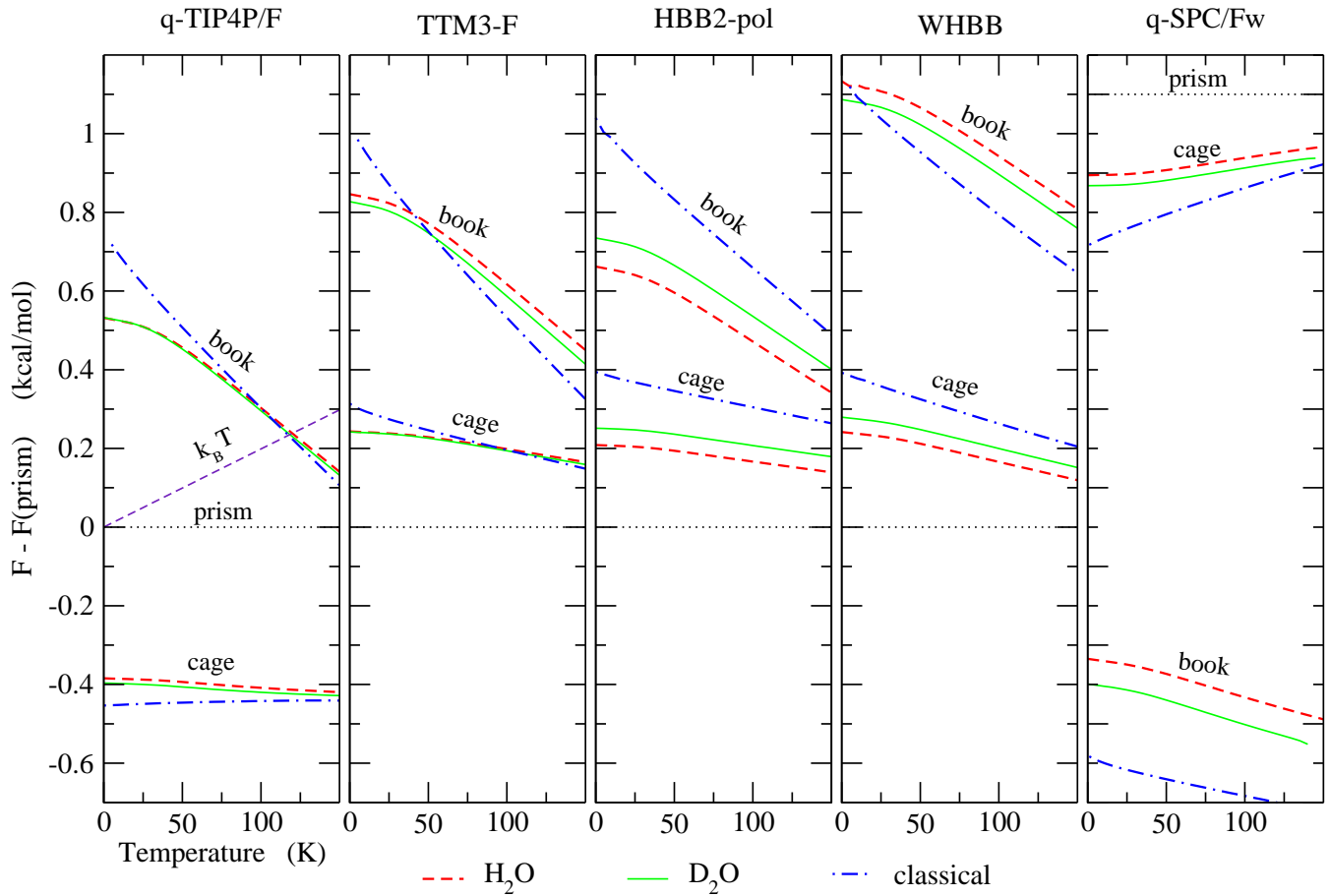}
	\end{centering}
	\caption{ \label{fig:all5} Free energy differences $\Delta F:=F-F_{\text{prism}}$ 
	of cage and book isomers of water hexamer  
with respect to the prism isomer, computed by SCP with rotational
corrections ({\it cf.} \Eq{eq:SCP-rot}) for five different water potentials 
for both classical (H$_2$O)$_6$ and quantum (H$_2$O)$_6$ and (D$_2$O)$_6$.  
Note that the y-axis for \mbox{q-SPC/Fw} is shifted, as indicated by the dotted line 
corresponding to the prism reference.} 
\end{figure*}

The quantities of interest are the free energies of various isomers with respect to that of the prism isomer, 
ie., the differences $\Delta F:=F-F_{\text{prism}}$ as a function of temperature $T$.  
\Fig{fig:all5} shows the results for $\Delta F$ for the cage and book isomers of classical (H$_2$O)$_6$ 
and quantum (H$_2$O)$_6$ and (D$_2$O)$_6$, for each of the five potentials specified above.  
The calculations were carried out following the protocol described in Ref.~\onlinecite{Brown2013}, 
with the addition of a rotational correction, described here in the appendix.

While the absolute free energy for each isomer (not shown) was found to change dramatically and in a predictable fashion, 
due to either isotopic substitution of hydrogen in particular or to quantum effects in general, 
the effects are much smaller for the free energy differences.  
For all five potentials both the isotope shift $\delta F_{D_2O}:=\Delta F_{\rm D_2O}-\Delta F_{\rm H_2O}$ 
and quantum shift $\delta F_{\hbar=0}:=\Delta F_{\hbar=0}-\Delta F_{\rm H_2O}$ are quite small compared to the $\Delta F$ values, 
ie., neither isotopic substitution or even going from classical water hexamer to quantum water hexamer would change 
the energy ordering of the prism, cage, and book isomers.  
A relatively small sensitivity of thermodynamic properties of water to isotopic substitution 
is though a well established fact both experimentally 
and theoretically 
(see, e.g., Refs.~\onlinecite{Habershon2009, Markland2012}). 
Arguably, the most striking feature of these results is the much smaller quantum and
isotope shifts for \mbox{q-TIP4P/F} and \mbox{TTM3-F} than for the other three potentials.
The effect is actually so small that it is hardly possible to reproduce accurately 
using a replica exchange path integral simulation due to statistical errors \cite{comment}.

While a thermodynamic integration method, specifically that adapted to the path integral Monte Carlo framework \cite{Ceriotti2013},
would seem to be better suited, as it is designed to compute the isotope shift directly, very long simulation times would 
still be needed in order to achieve a sufficiently small statistical error.  
The SCP method used in this work does not suffer from such convergence problems.  
Yet,  before any reliable path integral simulations become available, 
which may or may not confirm the present results, 
the manifestly approximate nature of SCP requires some justification and discussion.
In the absence of any reliable path integral results, etc., to support the (manifestly approximate) SCP results presented here, we consider the following two possibilities.  
First, in spite of its approximate nature, the SCP estimates of the free energy differences are actually accurate due to massive cancellations of the systematic errors.  
Second, the observed anomalies in the isotope effect are actually
artifacts of the inherent approximations of the SCP method.  
Since so far numerous numerical evidence is in favor of the SCP method
(e.g., Refs.~\onlinecite{Georgescu2013, comment}), in the following
discussion we disregard the latter possibility (unless proved to be the case) as being not constructive. 

Supposing the SCP results to be correct, the question is why the quantum and isotope effects for 
\mbox{q-TIP4P/F} and \mbox{TTM3-F} are substantially smaller than for the other three potentials.  
According to the analysis of Habershon {\it et~al.} \cite{Habershon2009}, 
the absence of anharmonic terms in the \mbox{q-SPC/Fw} description of the OH stretch 
could account for a larger isotope effect than that seen in \mbox{q-TIP4P/F}.  
The argument then is that it is the anharmonic terms in the OH stretch that are responsible for
cancellations of the competing quantum effects in the \mbox{q-TIP4P/F} water,
resulting in a small isotope effect.  
Since \mbox{TTM3-F}  is also highly anharmonic, this same argument can be used to explain the 
very small isotope and quantum effects observed with this PES. 
However, in both WHBB and \mbox{HBB2-pol}, the monomer contribution is based on the 
same Partridge-Schwenke fit \cite{PartridgeSchwenke1997}
 used in the \mbox{TTM3-F} model, yet the larger isotope/quantum shifts for WHBB and \mbox{HBB2-pol} are 
 comparable to those seen with \mbox{q-SPC/Fw}.  
As such, it appears that the relatively large isotope effects in both {\it ab~initio} potentials must be due to their
explicit inclusion of three-body terms.
Finally, we note again the constraint of permutational invariance, present
only in the parametrization of the {\it ab~initio} potentials, which, in principle, allows for hydrogens to
be transferred between oxygens.  
Neither \mbox{q-TIP4P/F} nor \mbox{TTM3-F} are permutationally invariant, 
i.e., each hydrogen remains assigned to its oxygen in these models.  
Still, the inclusion of this property does not guarantee that hydrogen 
exchange is accounted for accurately by either WHBB or \mbox{HBB2-pol}.  
In the context of this work, the key implication of this property is the much greater coupling via intermolecular 
degrees of freedom in the {\it ab initio} potentials.  
In other words, the quasi-separability of the intermolecular and intramolecular 
degrees of freedom present in \mbox{q-TIP4P/F} and \mbox{TTM3-F} models is much less pronounced in 
WHBB and \mbox{HBB2-pol}, leading to greater quantum effects associated with much
more flexible hydrogen degrees of freedom.

\section*{Acknowledgements}
This work was supported by the National Science Foundation (NSF) Grant No. \mbox{CHE-1152845}.  
SEB was partially supported by NSF Grant No. \mbox{DMS-1101578}. 
Volodymyr Babin and Francesco Paesani are acknowledged for discussing with 
us their results on water hexamer and for sending us the source code for the \mbox{HBB2-pol} PES.

\appendix 
\section{Free energy within the quasi-harmonic approximation with rotational correction.}

Consider an $N$-atom cluster and its isomer corresponding to a relatively deep and stable
potential energy minimum, i.e., we assume that it is separated from
the rest of the configuration space by relatively large energy barriers.  
In the absence of an external field the translations of the center of mass
can be separated so that we may consider the subspace $\mathbb{R}^{(3N-3)}$ that includes 
only the vibrational degrees of freedom and the rotations of the whole cluster.  
Because the potential energy $U(\mathbf{r})$ is invariant to these rotations,
the energy minimum is a ($3$)-dimensional manifold. 

To further simplify the problem consider a harmonic approximation or, more generally, a
quasi-harmonic approximation, represented by an effective (generally temperature-dependent) harmonic
Hamiltonian. To be concrete consider the quasi-harmonic approximation arising within the SCP
method \cite{Brown2013}, as used in this work. SCP considers
the system in the so called ``Eckart subspace'',
which is a reduced ($3N-6$)-dimensional subspace orthogonal to the translational
and rotational degrees of freedom (see, e.g., the discussion in
ref. \onlinecite{Brown2013}). Let $\hat H_{h}(T) $ define the (generally temperature-dependent) 
effective harmonic Hamiltonian, and $\omega_k$
($k=1,...,3N-6$), the corresponding efective harmonic frequencies.
The free energy in the Eckart subspace for a single minimum is then approximated by

\begin{multline}
F(T)\approx \< U \>_{h}  \\
+ \sum_k \Bigg[ \kbt \log \left(
 2\sinh \frac {\hbar\omega_k}{2\kbt} \right)
-\frac{\hbar \omega_k } 4 \coth \frac  {\hbar\omega_k}{2\kbt} \Bigg] \; ,
\label{eq:F1}
\end{multline} 
where $\< U \>_{h} $ defines the thermal average of the original potential 
over the reference harmonic system.

The above approximation to the free energy completely ignores the
rotational contribution, which may be important for small enough clusters.  
In order to include it we propose the use of a rigid asymmetric top correction 
\cite{McQuarrieBook2000}, which with the omission of terms that cancel when the 
energy difference between isomers is considered, reduces to 

\be
F_{\rm rot}(T)\approx -\frac {\kbt} 2 \log \frac {I_1 I_2 I_3(\kbt)^3}{\hbar^6} + \kbt \log\sigma  \; ,
\ee
where $I_1$, $I_2$ and $I_3$ are the principal moments of inertia of
the isomer evaluated at its minimum configuration, and $\sigma$ is the order of the isomer point group, 
which is unity unless the isomer configuration has symmetries.  
Consequently, the rotational correction to the free energy difference for two isomers 
$A$ and $B$ can be estimated using
\be \label{eq:SCP-rot}
\Delta F_{\rm rot} \approx \kbt \left[\frac 1 2 \log
\frac{I^B_1I^B_2I^B_3} {I^A_1I^A_2I^A_3}  + \log
\frac{\sigma^A}{\sigma^B}\right]\; .
\ee
Note that for water hexamer the rotational contribution to the
free energy difference between between different isomers is of the
order of $\sim 0.1 \kbt$.

\end{document}